# A new approach for rain gush formation associated with ionic wind

S. L. Chin
Center for Optics, Photonics and Laser (COPL)
Laval University
Quebec City, QC G1V 0A6
Canada.

Xueliang Guo[3], Huanbin Xu
Key Laboratory for Cloud Physics, Chinese Academy of Meteorological Sciences
Zhong-guan-cun South Avenue No.46, Haidian District, Beijing 100081, China

Fanao Kong, Andong Xia, Hongmei Zhao and Di Song
Institute of Chemistry
Chinese Academy of Sciences
Beijing, China

Tie-Jun Wang[1], Gengyu Li, Sheng-zhe Du, Jingjing Ju, Haiyi Sun, Jiansheng Liu, Ruxin Li[2], Zhizhan Xu
State Key Laboratory of High Field Laser Physics, Shanghai Institute of Optics and Fine Mechanics, Chinese Academy of Sciences, Shanghai, China.

Corresponding authors: tiejunwang@siom.ac.cn[1]; ruxinli@mail.shcnc.ac.cn[2]; guoxl@cma.gov.cn[3]

*Abstract*

Based upon experimental observation in the laboratory, we propose that ionic wind from corona discharge inside a thundercloud would play an important role in producing a rain gush. A cyclic chain of events inside a super-saturated environment in a thundercloud is proposed, each event enhancing the successive ones until lightning occurs. These successive events are collision between snowflakes and rimers, charge separation, corona discharge, avalanche ionization, ionic wind originating from the positively and negatively charged masses of cloud, vortex motion and turbulence when mixed with the updraft, more collision, more charge separation, stronger corona discharge, and so on. Meanwhile, avalanche ionization would produce more CCN (cloud condensation nuclei) resulting in more precipitation and hence rimers formation in the super-saturated environment. More collision in the buoyant turbulence would lead to more fusion of droplets and the formation of larger rimers. The cyclic processes would repeat themselves until the electric field between the two oppositely charged masses of cloud was strong enough to induce a breakdown. The latter would create a sudden short circuit between the two charged masses of cloud neutralizing the charges. There would be no more ionic wind, hence, much less buoyant turbulence.

The updraft alone would not be sufficiently strong to support larger rimers which would fall down 'suddenly' to the earth surface as a rain gush.

1. Introduction

It is well-known that after a strong lightning in a thundercloud with a strong updraft of moist air (convective cloud), a gush of rain will fall down with a good probability [1-6]. Several mechanisms have been proposed to explain how lightning can trigger the rain gush. Early scientists proposed that the charged particles were levitated in the strong thunderstorm electric fields until a lightning flash destroyed the field [1]. Then a numerical model [7] that couples the growth of the cloud particles with electrical development was developed, but failed to agree with the observations on variations of electric field and precipitation rate following lightning strokes [8,9]. Rain gush was also explained through the collisions and coalescence of cloud droplets under lightning electric field firstly proposed by Moore at al. [3]. Thunderstorm electrification could play an important role in the development of precipitation in the theory [10]. The above hypothesis is based on the fact that lightning occurs first and then triggers precipitation. In this hypothesis, the particles through electrification process must move some distance through the cloud before they can grow and become large enough to fall down to the ground as rain. This may take several minutes and longer. Aayaratne and Saunders [11] suggested an alternative hypothesis where the lightning flash is caused by the falling precipitation. They assumed the flash is initiated by the positively charged graupel pellets comprising the commonly observed lower positive charge center. The enhanced local electric field around this positive charge pocket may trigger a ground flash from the main negative charge center above. Then the following precipitation may be relatively close to the ground resulting in a rain gush within the observed short time intervals. This theory does not require a long time internal between the lightning flash and rain on the ground, which seems to agree with the observation of rain gush in 2-6 minutes after overhead lightning [5]. Another theory based on radial wind was also proposed [12,13]. The proposed radial wind could be generated by acoustic wave (explosions or lightning), leading to an increase on the rate of coalescence of water droplets and hence triggering precipitation. However, these theories are still under debate [5,10].

In a thundercloud, there are strong updraft and downdraft [14, 15]. Rain gush is believed to be due to a much stronger downward draft that pushes the existing raindrops and ice and snow to fall more rapidly [14]. However, such a downward draft was not identified physically. In such a thundercloud, there exist static strong electric fields (high voltage) and hence, there are a lot of corona discharges (leaders) and streamers development before intra or extra cloud lightning [16] occurs. Therefore, corona discharge induced ionic wind would occur before lightning. We believe that this ionic wind would play an important role to induce the rain gush. This latter explanation is based upon our laboratory observation and measurement of corona discharge induced ionic wind blowing around high voltage electrodes inside a cloud chamber and in air.

Ionic wind is a consequence of corona discharge in which avalanche ionization near a high voltage electrode occurs. Let's say an electrode is positively charged to a high voltage in air so that the electric field around it is strong enough to induce a corona discharge; i.e. avalanche ionization occurs near the electrode. During avalanche ionization, electrons will be accelerated towards and into the positively charged electrode. Meanwhile, the bunch of much heavier positively charged ions will be accelerated outward from the ionization zone near the electrode along the electric field lines towards surrounding grounds. The slower moving positive ions will drag along with them a parcel of neutral air molecules via collisions and momentum transfer. Together, they constitute the ionic wind. If the electrode is negatively charged, avalanche ionization near the electrode will still take place at a certain high voltage. In this case, the positively charged ions will be accelerated towards the electrode and are neutralized by taking up electrons from the electrodes. These neutral particles will be scattered around by the electrode giving rise to a local air motion. However, the electrons will be accelerated away from the electrode. They will encounter oxygen molecules ($O_2$) in air and the ozone molecules ($O_3$) created in the corona discharge. Most of these electrons will adhere to the oxygen and ozone molecules resulting in stable negatively charged ions because of their strong electron affinities of 0.45 eV and 2.10 eV [17], respectively. These negatively charged ions will be pushed away (accelerated) from the negatively charged electrode. Similar to the case of positive ions, they will drag along with them a parcel of air molecules through collision and momentum transfer giving rise to an ionic wind blowing away from the electrode.

Ionic wind has found applications on cooling device for LEDs [18] and aerodynamic actuators [19]. This ionic wind phenomenon is even successfully used as a propulsion technology in recent space missions [20, 21]. The ion propulsion technique was used to correct the trajectories of satellites [20] and spacecraft [21]. For example, the first mission of NASA's New Millennium Program, Deep Space 1, was propelled by ion thruster engine and launched on October 24, 1998 [20].

In this work, we shall give a description of the laboratory experiment on ionic wind followed by an explanation of our thought on the role played by ionic wind in triggering a rain gush.

## 2. Laboratory experiment

In the laboratory, different experiments were carried out. In the first experiment, two identical copper electrodes were set inside a diffusion cloud chamber (0.5 x 0.5 x 0.2 m$^3$). The axes of the electrodes were parallel to the bottom base plate and they were set roughly perpendicular to each other. Each electrode had a conic shape (1.1 cm (bottom diameter) × 1.6 cm (length) with the diameter of the tip ∼0.5 mm). One of them was grounded while the other, positively charged, connected to a high voltage power supply. The tips were about 2 cm apart and were set at a height of about 2 cm from the bottom cold plate. The temperature at the bottom well-grounded metallic cold plate was set at -46°C while an open water bath with pure water at 20°C was at 17.5 cm above the cold plate. The bottom part of the cloud chamber around the two electrodes was illuminated by an expanded green laser beam. The chamber was sealed thermally. Before applying the high voltage, the cloud chamber was left alone to 'settle down' for more than 30 minutes. Snowflakes and ice particles (to be collectively called snowflakes in what follows) accumulated on the bottom cold plate (Fig. 1a). The high voltage power supply was then turned on. The voltage was increased gradually from low to high. As explained in the introduction, corona discharge induced ionic wind started to blow from the positively charged electrode; it blew up the snowflakes from the cold plate making the motion of the snowflakes visible to the naked eyes [22].

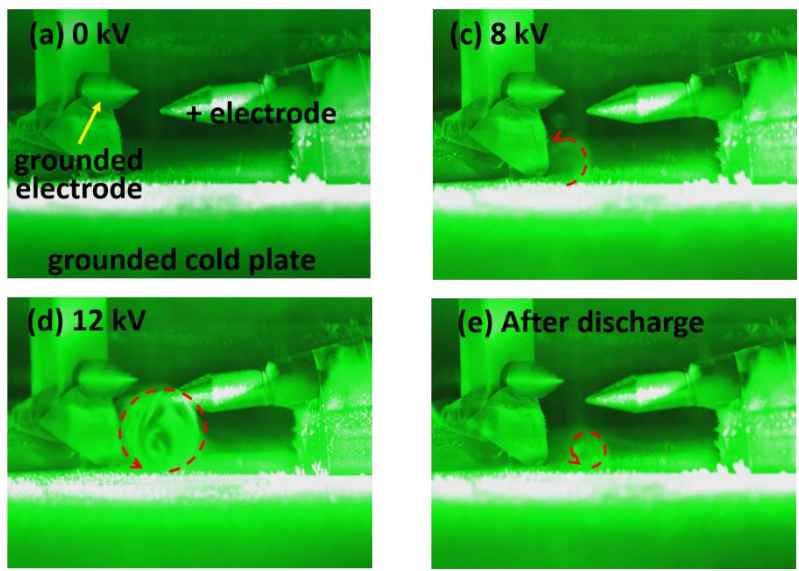

Figure 1. Snapshots of experimental environment inside a thermally sealed diffusion cloud chamber. Two pointed electrodes (separated by about 2 cm) with cylindrical bases were set roughly at right angle to each other. Their axes are roughly parallel to the bottom cloud plate whose temperature was set at −46°C. An open water bath (not shown) with pure water at 20°C was at 17.5 cm above the cold plate. The region around the electrodes was illuminated by a green laser. (a) The background after the cloud chamber was cooled down for more than 30 minutes; (b) high voltage

at 8 kV; (c) high voltage at 12 kV; (d) soon after a breakdown occurred between the two electrodes at a potential difference of slightly higher than 12 kV.

As the high voltage increased, the ionic wind speed increased. The ionic wind follows the electric field lines from the positive electrode to various places of the ground plate, including the grounded electrode. When the ionic wind hit the ground plate, it would turn into an anticlockwise turbulent vortex as indicated by the red curved arrow in Fig. 1b with the positive electrode biased at 8 kV. However, there was no electrical breakdown yet between the two electrodes until the high voltage reached 12 kV. Around this voltage, there was a visible surge in the strength of the ionic wind becoming a strong turbulent vortex indicated by the red circle in Fig. 1c. It was when the high voltage was increased slightly beyond 12 kV that a sudden electrical breakdown occurred between the tips of the two electrodes. Immediately, the corona discharge and ionic wind stopped (Fig. 1d). But the corona discharge and ionic wind grew again very quickly followed by another breakdown between the two electrodes, and so on. Video 1 shows a dynamic evolution of the corona discharge and ionic wind followed by breakdown at higher voltages.

When breakdown occurred, a highly conducting plasma channel was created through which electrons in the plasma would be accelerated towards the positive electrode. The charges on the positive electrode would be neutralized almost 'instantaneously'. Consequently, corona discharge and ionic wind would stop instantaneously. But since the positive electrode was always connected to the high voltage power supply, a rapid charge up would induce another series of corona discharge, ionic wind and breakdown, and so on. Because of the fast recovery of the corona discharge and breakdown, there would be some residual airflow between two breakdown events. That is shown in Fig. 1d where we can see a small turbulence blowing below the positively charged electrode when there was no breakdown.

We carried out further experiments in air inside a Faraday cage. The purpose was to measure the wind speed caused by the corona discharge in air. Since inside a thundercloud, both positively and negatively charged clouds are present and are separated from each other, we measured the ionic wind speed from both positively and negatively charged electrodes. One of the identical electrodes used in the cloud chamber was used in this experiment. It was fixed in air with the axis parallel to the horizontal plane at a height of 19.5cm from the bottom grounded metallic plate. The distance from the tip to the four grounded walls of the Faraday cage and the grounded top cover was about 15cm. The electrode was charged either positively or negatively. Ionic wind speed was measured as a function of the applied voltage. Calcium carbonate (CaCO$_3$) powder whose averaged size was roughly 20-100 µm was spread by hand into the Faraday cage near the electrode while an expanded green CW laser beam illuminated the region right below the electrode's tip. The powder was blown away by the ionic wind. Using a high-speed digital video camera filming at 60 frames per second, successive frames of the movement of a calcium carbonate particle were captured. The distance travelled by a particle between two successive frames (the length of a trace in the picture) together with the frame speed would give the speed of the particle. For each applied voltage, we selected the highest speed. We approximated this speed as the ionic wind speed keeping in mind that the real wind speed should be higher.

Fig. 2a shows a picture of the calcium carbonate powder blown away from the electrode at +45 kV in the case of a positive corona (electrode charged positively). Video 2 shows in more detail the blowing of the powder as the applied voltage increased.

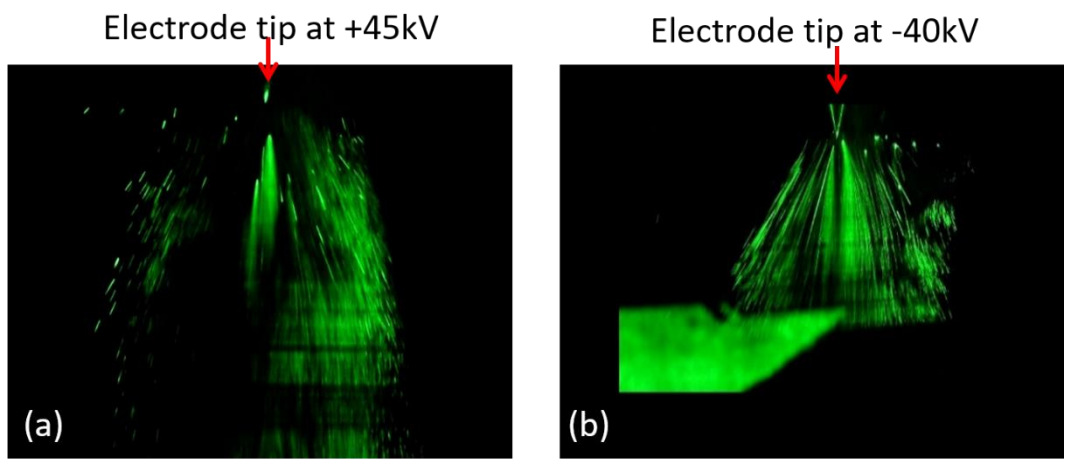

Fig. 2. Calcium carbonate powder blown away from the positively (a) and negatively (b) charged electrodes.

In the case of a negative corona (electrode charged negatively), the ionic wind still blew away from the negatively charged electrode as expected in the discussion in the introduction due to electron attachment onto O$_2$ and O$_3$. This is shown in video 3. Fig. 2b shows a picture of the calcium carbonate powder blown away from the electrode at -40 kV. Fig. 3 shows the plots of the ionic wind speed as a function of the applied voltage in both cases of positive and negative corona discharges. Beyond the highest voltage of about 50 kV, breakdown occurred and there was no more ionic wind.

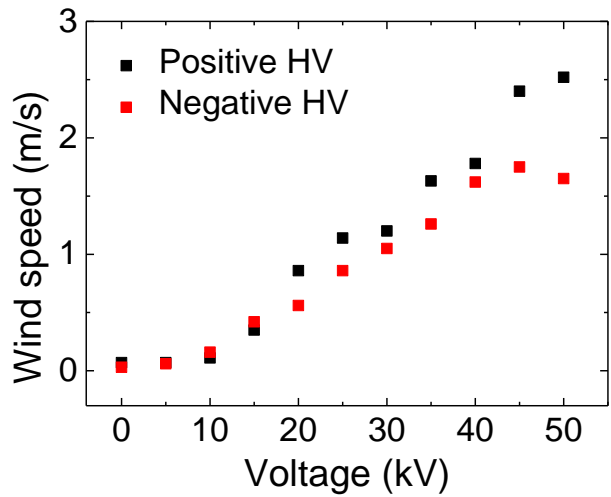

Fig. 3. Measured maximum ionic wind speed from positive and negative coronas as a function of the applied voltage. At -45 kV and -50 kV in the case of a negative corona, there was some leakage in the electrical circuit rendering the wind speed slightly lower.

We note that the ionic wind speed is of the order of a few m/sec before breakdown occurs. This wind speed is similar in magnitude to the wind speed inside a thundercloud [15]. This means that the ionic wind before breakdown would have a force strong enough to influence the updraft in a thundercloud.

We did another experiment in which two identical oppositely charged electrodes (same electrodes used in the previous experiments) were supported in air with their axes parallel to the horizontal plane at the central region of the Faraday cage. The tips of the electrodes were separated by 6 cm. Mosquito incense was burned inside the Faraday cage and an expanded green laser beam illuminated the smoke. When a high voltage was applied, vortices and turbulence of the air between the electrodes were observed. Fig. 4 shows an example of the air motion when the applied voltage was 26 kV between the electrodes. Video 4 gives a more detailed view of the rotating air motion.

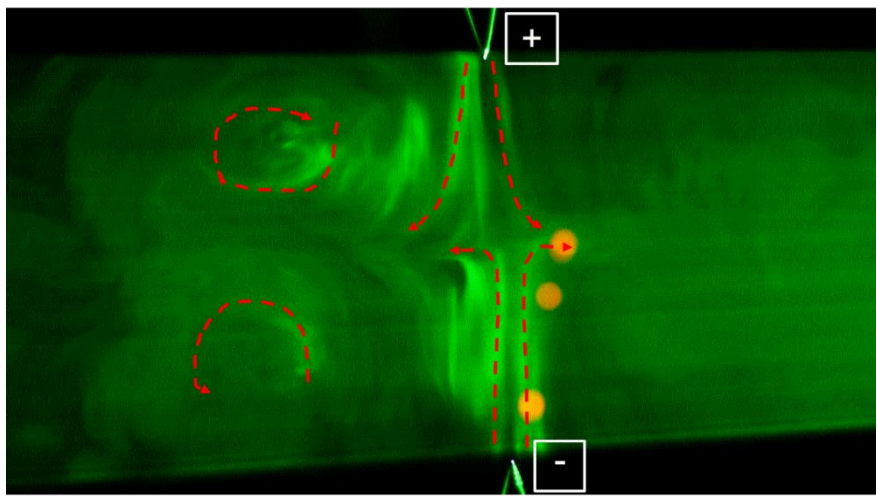

Fig. 4. Top view of vortices and turbulence in air induced by the ionic winds from two oppositely charged electrodes set in the horizontal plane. The arrows indicate the direction of vortex motion. The distance between the tips of the electrodes was 6 cm. The voltage between them was 26 kV. The three orange 'balls' were the images of the burning tips of the mosquito incense at the bottom of the Faraday cage.

A corona discharge and the associated ionic wind in a thundercloud should be similar to those observed in the laboratory experiment described above because they are similar electrostatic phenomena. For example, in the laboratory experiment in a cloud chamber, the environment around the electrodes near the cold plate in the cloud chamber contained ice particles, moisture, small droplets of water, etc. It is similar to the environment in a thunder cloud. The electric field strength required to induce avalanche ionization in a corona discharge would be similar in the two cases. In fact, in a thundercloud, the electric field could be up to several hundred kV/m [23]. This is similar to the electric field between the two electrodes in our first experiment just before breakdown which was ~ 12 kV/2cm = 600 kV/m. Thus, our experimental observation could be applied to illustrate the electrostatic phenomena occurring in a thundercloud.

3. Rain gush

It is well-known that positive and negative charges are separated continuously inside a thundercloud where there is super-saturation [15]. Very often, the top part of the cloud containing mostly smaller ice particles (ice crystals or snowflakes) is charged positively while the bottom part containing mostly rimers (graupel particles or hailstones, etc.) is charged negatively [15]. We shall concentrate upon this type of cloud. Fig. 5 gives a schematic drawing of the cloud structure adopted from ref. 15. In the main super-saturated charging zone, positive and negative charges are separated continuously; the detailed mechanism is still a question of debate [14, 15]; we shall not go into it.

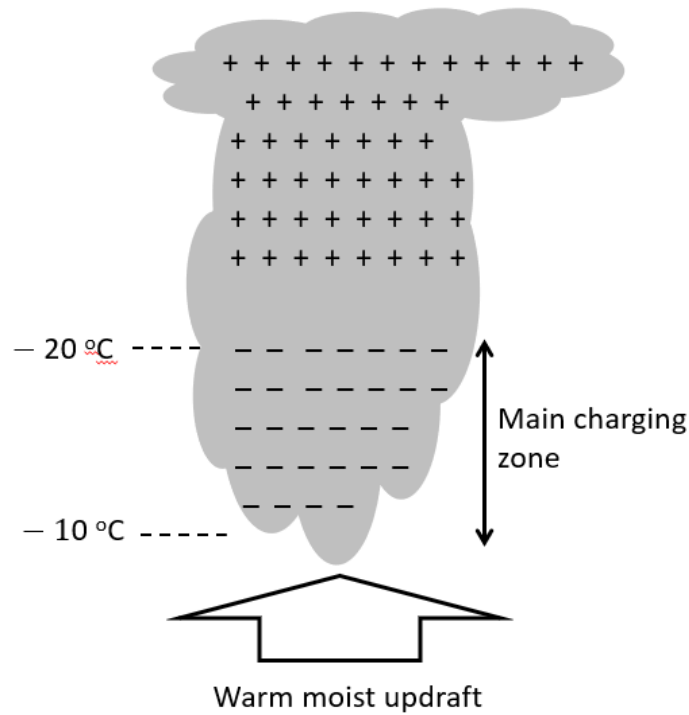

Figure 5. A schematic drawing of a thundercloud structure adopted from ref. 15, p. 252 – 254.

According to the current understanding, when the ice crystals or snowflakes collide with the rimers (graupels etc.), charge separation occurs [15]. The lighter snowflakes are charged positively and move upward forming the upper part of the cloud. The heavier rimers are charged negatively and fall downward. They are suspended by the updraft at the bottom part of the cloud (see Fig. 5). For the sake of clarity in the discussion, we shall separate the main charging zone into two conceptual zones, a mixing zone where charges are separated and a bottom zone where rimers are suspended by the updraft. This is shown in Fig. 6. The upper and lower charged clouds are now equivalent to two oppositely charged electrodes. The supersaturated mixing zone contains super-cooled water droplets, ice particles, snowflake, rimers, etc. [15]. We further assume that the updraft would supply a sufficient amount of moisture into the mixing zone continuously so that super-saturation is always maintained.

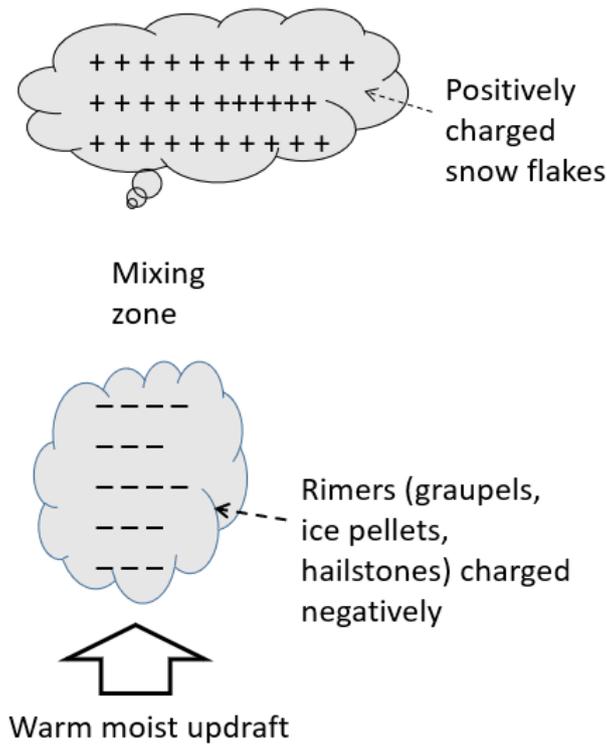

Figure 6. Schematic of a thundercloud structure with the main charging zone separated conceptually into two zones, a mixing zone where charges are separated and a bottom zone where rimers are suspended by the updraft. Super-saturation in the mixing zone is assumed.

When more and more charges were separated, the cloud would be highly charged positively at the top part and negatively at the bottom part of the cloud with the super-saturated mixing zone in between as shown in Fig. 6. The charge distribution in both the upper and the lower parts of the cloud would be random. Some zones would have a higher density of charges while some other zones less. We assume that the charge density fluctuation in the cloud was small. According to our experimental results in the laboratory, we would expect that strong ionic wind would mostly blow outward from both the top (positively charged) and bottom (negatively charged) parts of the cloud along the general directions of the electric field lines between them. The ionic wind is indicated in Fig. 7 by the dashed arrows in a symbolic way. The ionic wind speed would be a few m/sec according to our laboratory measurement (Fig. 3). This speed is similar in strength to that of the updraft [15, 24, 25]. This updraft would blow into the mixing zone as indicated by the broad dotted arrow in Fig. 7 and mix with the ionic winds.

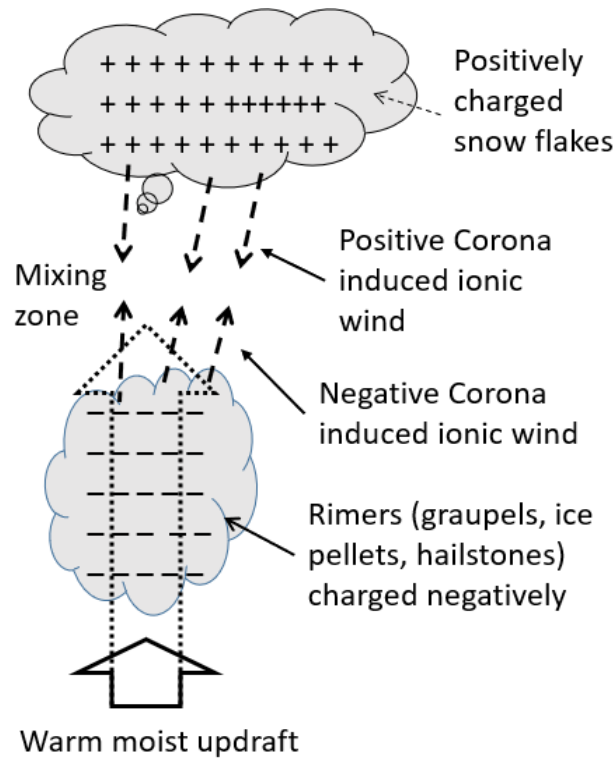

Fig. 7. Schematic drawing showing the ionic winds (dashed arrows) blowing both upward and downward from the bottom and top part of the cloud. Super-saturation is assumed in the mixing zone. The updraft blows into the mixing zone in reality. This is indicated by the broad dotted arrow. The speed of the updraft and that of the ionic wind from the corona discharges are comparable, of the order of a few m/sec. They would blow into one another forming turbulence and vortex motion in the mixing zone.

The consequence of the counter blowing ionic wind together with the updraft would be the formation of vortices and turbulence in the mixing zone as our experiment showed (Fig. 4). This would result in churning and rotating buoyant air masses containing a 'soup' of rimers, snowflakes, ice particles, super-cooled water droplets, supersaturated air, etc. inside the mixing supersaturated zone. A series of cyclic enhancing events would follow (Fig. 8). For example, avalanche ionization in the corona discharges (box 4) would produce CCN (cloud condensation nuclei) (box 8) which would enhance precipitation and rimer formation [26, 27]. The churning and rotation of the air masses would increase the collision among the particles inside the mixing zone (box 1) and enhance the fusion of droplets, etc. Rimers would grow in size (box 7). Meanwhile, more collision between the snowflakes and the rimers etc. (box 1) would enhance charge separation (box 2) whereby more positively charged snowflakes would fly into the upper part of the cloud. More charge separation would enhance the electric field between the upper and lower parts of the cloud (box 3). This would increase the strength of the corona discharges (box 4). The strength of the ionic wind would also be increased (box 5). Stronger ionic wind would induce stronger turbulence and vortices (box 9) which would enhance stronger mixing (box 6), hence more collision (box1). More charge separation (box 2) would occur; and so on. In Fig. 8 showing the schematic of cyclic enhancement of the various events, boxes 1 to 6 constitute a closed cycle of events leading eventually to a breakdown while boxes 7 and 8 give rise to two 'leaky' processes, namely, enhanced precipitation and the growth of rimers during each cycle. Box 9 is the supporting 'forces' from the updraft and ionic winds providing humidity and buoyant vortex motion and turbulence to the churning air mass in the mixing zone.

The cyclic enhancement of the various events would come to an abrupt end when there were so many charge separations that a breakdown (lightning) 'suddenly' occurred between the top and bottom part of the cloud short-circuiting and neutralizing most, if not all, of the charges between the top and the bottom. Ionic wind would stop because there was no more (or very little) charges left in the cloud. (Justified by the first experimental observation. See Fig. 1d.) There would be no more turbulence and vortices to help suspending the 'soup' while the updraft would not be strong enough to suspend the large size rimers. A large quantity of rimers would thus suddenly fall down towards the ground resulting in a rain gush.

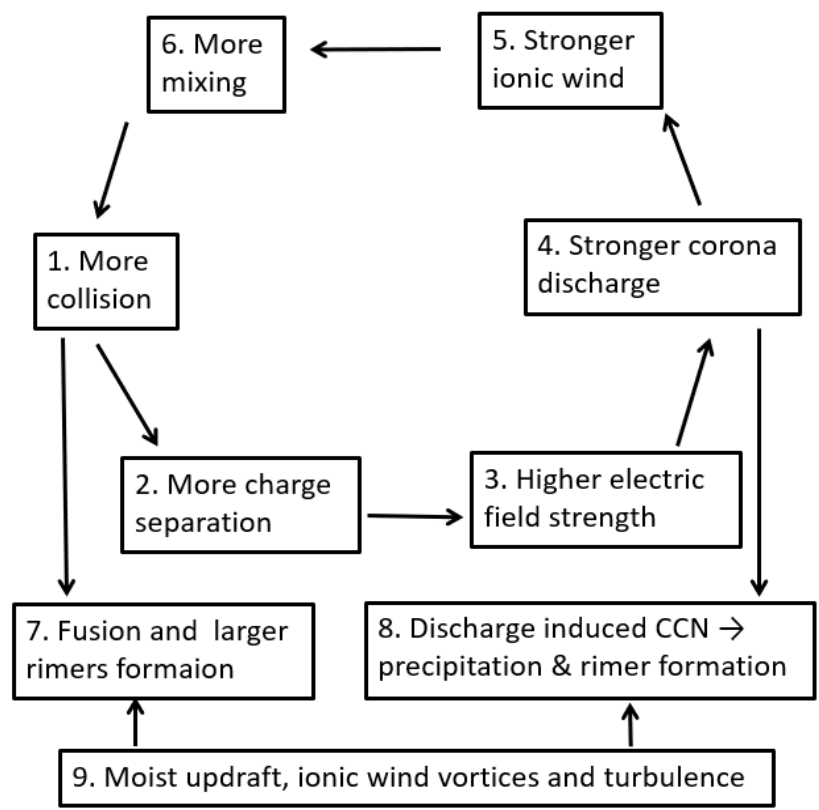

Fig. 8. Cyclic enhancement of various events in a thunder cloud where mixing of various particles (rimers, snowflakes, ice particles, super-cooled water droplets, etc.) in a super-saturated air would occur. It is assumed that the updraft continuously supply a lot of moisture into the cloud to maintain super-saturation. With more mixing, there would be more collision (box 1). More collision would lead to more fusion of droplets and larger rimers formation (box 7). They would be suspended by the buoyant turbulence due to the mixing of the updraft and the ionic wind (box 9). It would also lead to more charge separation (box 2) which would result in higher electric field strength between the top and bottom parts of the cloud (box 3). This would lead to a stronger corona discharge (box 4) (avalanche ionization). More CCN (cloud condensation nuclei) would be produced which would enhance precipitation and the formation of rimers (box 8) supported by the updraft and ionic wind (box 9). Stronger corona discharge would induce stronger ionic wind (box 5) and stronger churning of the air mass resulting in more mixing (box 6). More mixing means more collision (box 1); and so on.

The above cyclic mechanism could be applied in the case in which the top part of the cloud is charged negatively and the bottom part positively. This is because the ionic wind from the top and bottom part of the cloud would blow outward towards each other similar to the illustration in Fig. 7. If all other conditions were the same as in the previous case (top positive, bottom negative), the 'leaky' cyclic mechanism proposed in Fig. 8 would also be valid to trigger a rain gush.

4. Discussion

The cyclic chain of enhancement of the various events leading to a rain gush (Fig. 8) could be broken during its development. For example, if the amount of moisture were not sufficient in the updraft, the events in boxes 7 and 8 would not produce a sufficient amount of large rimers while the cyclic chain of events from box 1 to 6 would still go on. When a sudden breakdown (lightning) occurred, there would be little precipitation. In the extreme case, there would be no precipitation but only lightning. If the supply of moisture were not continuous, the amount of moisture in the mixing zone would fluctuate. The size of rimers would be smaller and their formation would be insufficient (box 7 and 8 in Fig. 8). However, the charge separation would still go on around the cycle from box 1 to 6. Eventually, a sudden breakdown would result in a weak rain gush or simply rain. If the updraft were not sufficiently strong, it would not be able to create sufficient buoyancy together with the ionic winds to help support large rimers in the 'soap'. Small rimers would fall down continuously while the events from box 1 to 6 would still go on. There would thus be no rain gush but lightning could still occur.

Thus, the most important factor to maintain the cyclic enhancement processes in Fig. 8 seems to be the continuous supply of sufficient moisture into the cloud by the updraft. That is to say, a rain gush would require at least the simultaneous satisfaction of the following important conditions, namely, a) sufficient amount of moisture in the updraft so as to maintain a super-saturation condition inside the mixing zone, b) continuous supply of moisture from the updraft and

c) sufficiently strong updraft. The probability for these conditions to be valid simultaneously would not be 100%. Thus, not all lightning strokes from a thundercloud would give rise to rain gush. We could try to make a very primitive estimation of this probability.

We assume that a convective thundercloud was formed and that the leaky cyclic events shown in Fig. 8 has taken place. Lightning would occur. For a rain gush to occur soon after a lightning stroke, we assume that the three conditions mentioned above would be satisfied. The strength of the updraft and the moisture content inside the updraft are independent in general. If we assume that once a continuous updraft were formed, the moisture content would not vary a lot, we would be left with two independent events to be satisfied, namely strong updraft and high moisture content. The updraft would have two possibilities, either strong enough or not strong enough. Similarly, the moisture content would have two possibilities, either high enough or not high enough to sustain super-saturation. Consequently, each of them (strong updraft and high moisture content) would have a probability of occurrence of 1/2. Together, the probability of having a simultaneous strong updraft and high moisture content would be ½ X ½ = ¼ or 25%. This is an upper limit. In reality, this probability should be lower because of the variation of the moisture content, the continuity of the updraft, etc. According to weather scientists, about 20% of lightning would result in rain gush (heavy rainfall) from a typical thundercloud [6]. This value is similar to the very primitive estimation of a probability of less than 25%.

Special cases of charge density fluctuation could exist. One would be that a certain zone at the top of the upper part of the thundercloud would have an especially high density of charges. This would result in corona discharge and ionic wind followed eventually by a lightning into the upper space. Similarly, the bottom part of the cloud could have an especially high charge density resulting in corona discharge and ionic wind blowing towards the ground. A lightning might eventually occur between the bottom part and the ground. These events would not give rise to rain gush.

5. Summary and conclusion

Based upon laboratory observations of ionic wind associated with corona discharge, we seem to be able to give a qualitative explanation of rain gush from a thundercloud. The main mechanism is proposed to be a cyclic chain of events inside a super-saturated environment (thunder cloud), each event enhancing the successive ones until lightning occurs. One could start from the collision between snowflakes and rimers which would lead to more charge separation which in turn would result in a stronger corona discharge with a stronger avalanche ionization. The latter would produce more CCN resulting in more precipitation and hence rimers formation in the super-saturated environment. The associated ionic wind originated from the positively and negatively charged masses of cloud would blow against each other while the updraft would mix with them. Together, they would result in vortex motion and turbulence. The latter would enhance the buoyancy of the air mass in the mixing or charge separation zone. It would also enhance the mixing of particles (snowflakes, rimers, super-cooled droplets of water, etc.). The latter would lead to more collision among them. More collision would mean more fusion of droplets and the formation of larger rimers in the mixing (charging) zone while more charges were separated. And the process would repeat itself until the electric field between the two oppositely charged clouds was strong enough to induce a breakdown. The latter would create a sudden short circuit between the two charged clouds neutralizing the charges. There would be no more ionic wind and larger rimers would fall down 'suddenly' to the earth surface as a rain gush.

## Acknowledgement

SLC acknowledges the support of Laval University, Quebec City, Canada. TJW acknowledges the supports from the Strategic Priority Research Program of the Chinese Academy of Sciences (Grant No. XDB160104), Key Project from Bureau of International Cooperation Chinese Academy of Sciences (Grant No. 181231KYSB20160045) and 100 Talents Program of Chinese Academy of Sciences, China.

Figure captions

Figure 1. Snapshots of experimental environment inside a thermally sealed diffusion cloud chamber. Two pointed electrodes with cylindrical bases were set roughly at right angle to each other. Their axes are roughly parallel to the bottom cloud plate whose temperature was set at −46°C. An open water bath (not shown) with pure water at 20°C was at 17.5 cm above the cold plate. The region around the electrodes was illuminated by a green laser. (a) The background after the cloud chamber was cooled down for more than 30 minutes; (b) high voltage at 8 kV; (c) high voltage at 12 kV; (d) soon after a breakdown occurred between the two electrodes at a potential difference of slightly higher than 12 kV.

Fig. 2. Calcium carbonate powder blown away from the positively (a) and negatively (b) charged electrodes.

Fig. 3. Measured maximum ionic wind speed from positive and negative coronas as a function of the applied voltage. At -45 kV and -50 kV in the case of a negative corona, there was some leakage in the electrical circuit rendering the wind speed slightly lower.

Fig. 4. Top view of vortices and turbulence in air induced by the ionic winds from two oppositely charged electrodes set in the horizontal plane. The arrows indicate the direction of vortex motion. The distance between the tips of the electrodes was 6 cm. The voltage between them was 26 kV. The three orange 'balls' were the images of the burning tips of the mosquito incense at the bottom of the Faraday cage.

Figure 5. A schematic drawing of a thundercloud structure adopted from ref. 15, p. 252 – 254.

Figure 6. Schematic of a thundercloud structure with the main charging zone separated conceptually into two zones, a mixing zone where charges are separated and a bottom zone where rimers are suspended by the updraft. Super-saturation in the mixing zone is assumed.

Fig. 7. Schematic drawing showing the ionic wind blowing both upward and downward from the bottom and top part of the cloud. Super-saturation is assumed in the mixing zone.

Fig. 8. Cyclic enhancement of various events in a thunder cloud where mixing of various particles (rimers, snowflakes, ice particles, super-cooled water droplets, etc.) in a super-saturated air would occur. It is assumed that the updraft continuously supply a lot of moisture into the cloud to maintain super-saturation. With more mixing, there would be more collision (box 1). More collision would lead to more fusion of droplets and larger rimers formation (box 7). They would be supported by the updraft and the vortices and turbulence (box 9). It would also lead to more charge separation (box 2) which would result in higher electric field strength between the top and bottom parts of the cloud (box 3). This would lead to a stronger corona discharge (box 4) (avalanche ionization). More CCN would be produced which would enhance precipitation and the formation of rimers (box 8) supported by the updraft and ionic wind (box 9). Stronger corona discharge would induce stronger ionic wind (box 5) and stronger churning of the air mass resulting in more mixing (box 6). More mixing means more collision (box 1); and so on.